\documentclass[a4paper, times, 10pt,twocolumn]{article}
\usepackage[top=4.9cm,bottom=3.7cm,left=1.5cm,right=1.5cm]{geometry}
\usepackage{ICMLC}
\usepackage{times}
\usepackage{graphicx}
\usepackage{indentfirst}
\usepackage{latexsym}
\usepackage[margin=8pt,font=footnotesize,labelfont=bf,labelsep=period
]{caption}
\usepackage{amsmath}
\usepackage{newtxtext}
\usepackage[varg]{newtxmath}
\usepackage{algorithmic}
\usepackage{algorithm}
\usepackage{cite}
\usepackage{subcaption}

\pdfpagewidth=\paperwidth
\pdfpageheight=\paperheight
\begin{document}

\title{An Encryption Method of ConvMixer Models without Performance Degradation}  

\author{\bf{\normalsize{Ryota Iijima${^1}$, Hitoshi Kiya${^2}$}}\\ 
\\
\normalsize{$^1$Department of Computer Science, Tokyo Metropolitan University, Hino, Tokyo, Japan}\\
\normalsize{$^2$Department of Computer Science, Tokyo Metropolitan University, Hino, Tokyo, Japan} \\
\normalsize{E-MAIL: iijima-ryota@ed.tmu.ac.jp, hitoshi kiya@tmu.ac.jp }\\
\\}

\maketitle \thispagestyle{empty}

\begin{abstract}
   {In this paper, we propose an encryption method for ConvMixer models  with a secret key.
   Encryption methods for DNN models have been studied to achieve adversarial defense, model protection and privacy-preserving   image classification.
   However, the use of conventional encryption methods degrades the performance of models compared with that of plain models.
   Accordingly, we propose  a novel method for encrypting ConvMixer   models.
   The method is carried out on the basis of an embedding architecture that ConvMixer has, and models encrypted with the method can have the same performance as models trained with plain images only when using test images encrypted with a secret key.
   In addition, the proposed method does not require any specially   prepared data for model training or network modification.
   In an experiment, the effectiveness of the proposed method is evaluated in terms of classification accuracy and model protection in an image classification task on the CIFAR10 dataset.}
\end{abstract}

\section{Introduction}
Deep neural network (DNN) models have been deployed in many applications including security-critical ones such as biometric authentication, automatic driving, and medical image analysis \cite{lecun2015deep, liu2019recent}.
However, they have been exposed to various threats such as adversarial examples, unauthorized access, and data leaks.
Accordingly, it has been challenging to train/test a machine learning (ML) model with   encrypted images as one way for solving these issues \cite{kiya2022overview}.
However, conventional methods that use models trained with encrypted images have caused performance to degrade compared with models trained with plain images. \par
Accordingly, in this paper, we propose a novel method based on a unique feature of ConvMixer \cite{trockman2022patches}, and it can overcome the above problems.
In the method, a model trained with plain images is encrypted with a secret key.
Also, to adapt to model encryption, test images are transformed with the same key.
The proposed method allows us not only to obtain the same performance
as models trained with plain images but to also update the secret key easily.
In an experiment, the effectiveness of the proposed method is evaluated in terms of performance degradation and model protection performance in an image classification task on the CIFAR-10 dataset.

\section{Related Work}
Conventional methods for encrypting DNN models and ConvMixer are summarized here.

\subsection{Model Encryption with Secret Key}
Many model encryption methods have been studied to be applied to adversarial defense, model protection \cite{aprilpyone2021block, maung2020encryption, maung2021ensemble, maungmaung2021a, maung2022privacy} and privacy-preserving image classification \cite{kiya2022overview, kawamura2020aprivacy, bandoh2020distributed, ibuki2016unitary, maekawa2019privacy, koki2020block, warit2021agan}.
Almost all model encryption methods are carried out by training models with images encrypted with a secret key, but the methods can degrade the performance of the models compared with that when using non-encrypted models due to the influence of encryption.  \par
Model encryption methods have to satisfy two requirements.
The first requirement is that authorized users with a secret key can obtain almost the same performance as that of non-encrypted models from encrypted models.
The second is that performance of the encrypted models is not high for unauthorized users without a correct key.
The proposed method aims not only to avoid the influence of encryption but to also provide an extremely degraded accuracy to unauthorized users.

\subsection{ConvMixer}
ConvMixer \cite{trockman2022patches} is well-known to have a high performance in image classification tasks, even though it has a small number of model parameters.
ConvMixer is a type of isotropic network.
It is inspired by the vision transformer (ViT) \cite{Alexey2021an}, so the architecture has a unique feature, called patch embedding. \par
\begin{figure*}[t]
    \centering
    \vspace{-10mm}
    \includegraphics[bb=0 0 960 313, scale=0.45]{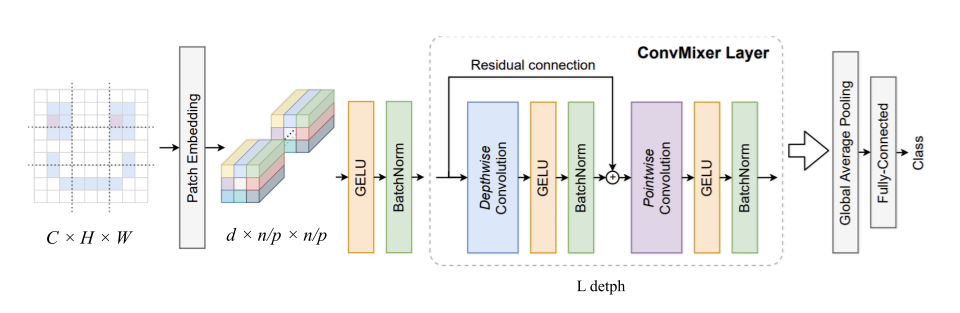}
    \vspace{-5mm}
    \caption{Architecture of ConvMixer \cite{trockman2022patches}}
    \label{fig1}
\end{figure*}

Figure \ref{fig1} shows the architecture of the network, which consists of two main structures: patch embedding and ConvMixer layer.
First, an input image $x \in \mathbb{R}^{C \times H \times W}$ is replaced with $z_0$ by patch embedding with patch size $P$ and embedding dimension $d$ as
\begin{equation}
    \begin{aligned}
    z_0 & = \mathrm{BN} (\mathrm{\sigma} \{ \mathrm{Conv}_{C \rightarrow d} ( x, \mathrm{kernel\_size} = P, \mathrm{stride} = P ) \} ). 
    \end{aligned}
\end{equation}
where $H$, $W$, and $C$ are the height, width, and the number of channels of $x$.
Also, $\mathrm{Conv}_{C \rightarrow d}$ is a convolution operation with $C$ input channels and $d$ output channels, $\mathrm{BN}$ is a batch normalization operation, and $\sigma$ is an activation function.
In addition, to simplify the discussion, we assume that $H$ and $W$ are divisible by $P$.
Next, $z_0$ is transformed into $z_l, l \in \{ 1, 2, ..., L \}$ by using $L$ ConvMixer layers.
Each layer consists of depthwise convolution $( \mathrm{ConvDepthwise} )$ and pointwise convolution $( \mathrm{ConvPointwise} )$ as follows.
\begin{equation}
    \begin{aligned}
    z'_l & = \mathrm{BN} ( \mathrm{\sigma} \{ \mathrm{ConvDepthwise} ( z_{l - 1} ) \}) + z_{l - 1} \\
    z_l & = \mathrm{BN} ( \mathrm{\sigma} \{ \mathrm{ConvPointwise} ( z'_l ) \})
    \end{aligned}
\end{equation}
Finally, the output of the $L$ th ConvMixer layer is transformed by Global Average Pooling and a softmax function to obtain a result. \par
In this paper, we utilize patch embedding in Eq.(1) to encrypt a model.
Patch embedding can be done in two steps.
\begin{enumerate}
    \setlength{\itemsep}{0cm}
    \item Reshape an input image $x$ into a sequence of flattened $2D$ patches ${x}_{\mathrm{p}} \in \mathbb{R}^{N \times (P^2C)}$, where $N=HW/P^2$ is the number of patches.
    \item Map each patch $\boldsymbol{x}^i_\mathrm{p} \in \mathbb{R}^{P^2C}$ to $\boldsymbol{z}^i_0$ with dimensions of $d$ as
    \begin{equation}
        \begin{aligned}
        \boldsymbol{z}^i_0 &= \boldsymbol{x}^i_\mathrm{p} \mathbf{E} \\
        \mathbf{E} & \in \mathbb{R}^{(P^2 C) \times d}.
        \end{aligned}
    \end{equation}
\end{enumerate}
A kernel in $\mathrm{Conv}_{C \rightarrow d}$ in Eq.(1) corresponds to $\mathbf{E}$ in Eq.(3).
In this paper, we show that model encryption can be carried out by transforming $\mathbf{E}$ with a secret key.
Also, this encryption does not degrade the performance of ConvMixer.


\section{Proposed Encryption Method}
Both a novel method for encrypting models and images, and the combined use of encrypted models and images are proposed here.

\subsection{Overview}
\begin{figure*}[t]
    \centering
    \vspace{-5mm}
    \includegraphics[scale=0.27]{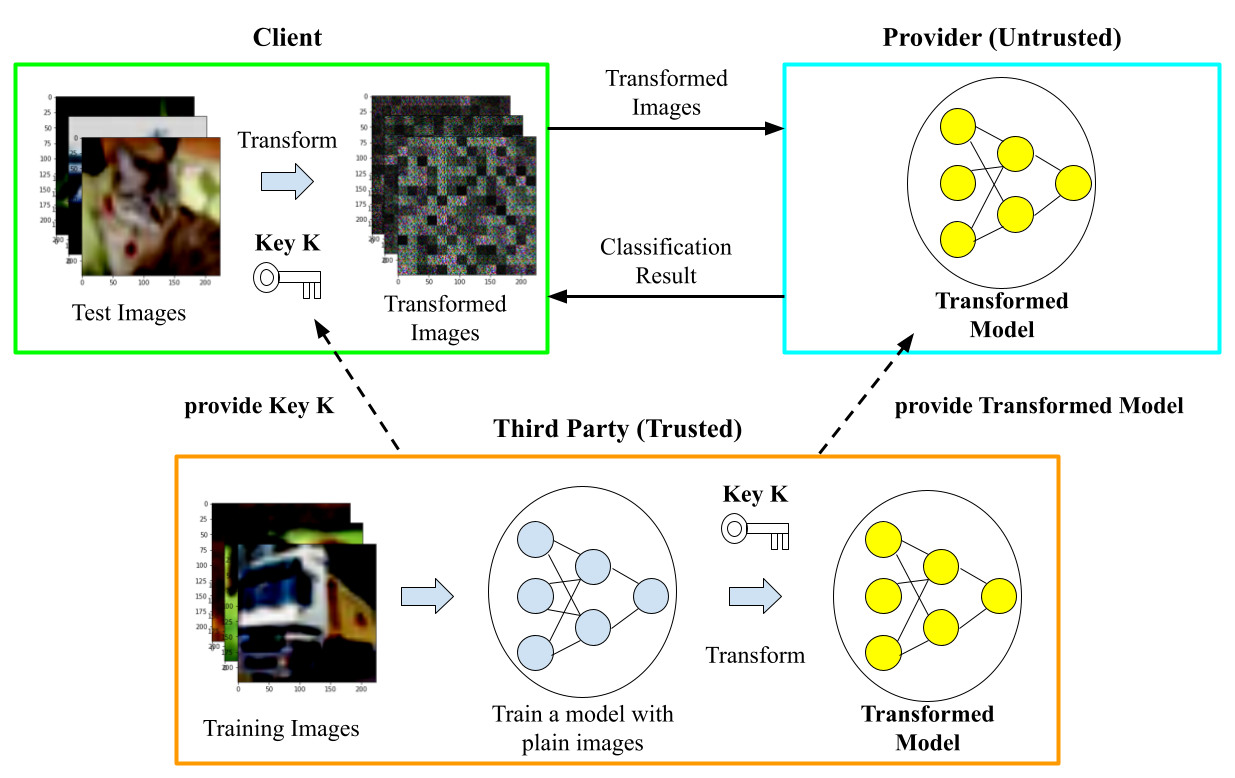}
    \caption{Scenario of proposed method}
    \label{fig2}
\end{figure*}

Figure \ref{fig2} shows an overview of the proposed method, where it is assumed that the third party is trusted, and the provider is untrusted.
The third party trains a model by using plain images and transforms the trained model with a secret key.
The transformed model is given to the provider, and the key is sent to a
client.
The client prepares a transformed test image with the key and sends it to the provider.
The provider applies it to the transformed model to obtain a classification result, and the result is sent back to the client.
Note that the provider has neither a key nor plain images.
The proposed method enables us to achieve this without any performance degradation compared with the use of plain images.

\begin{figure*}[t]
    \centering
    \includegraphics[scale=0.27]{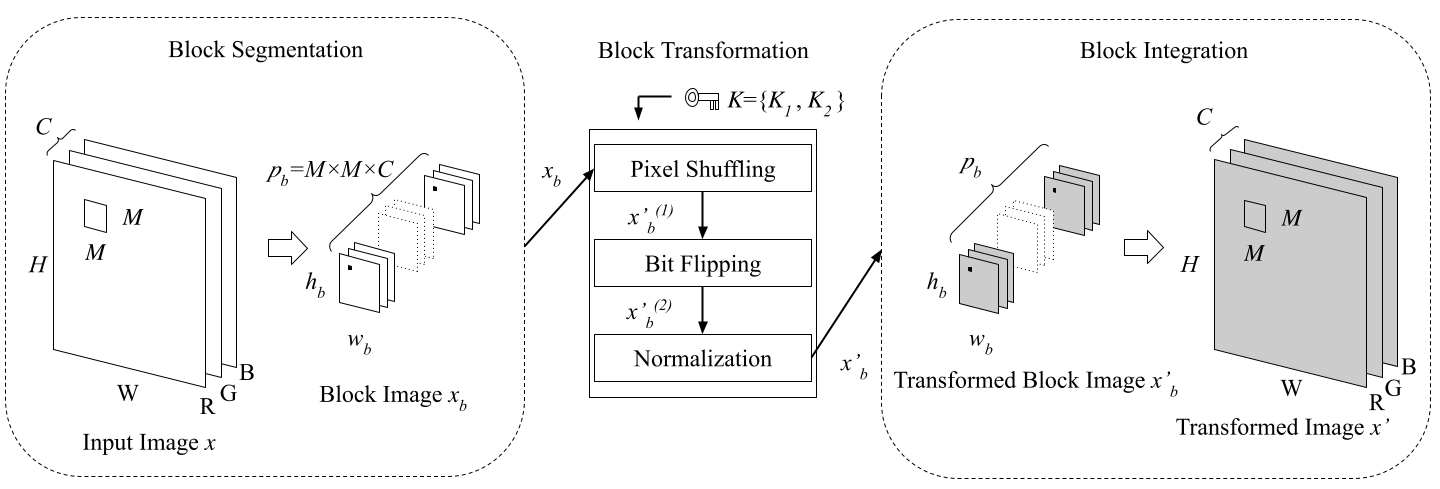}
    \vspace{-3mm}
    \caption{Procedure of block-wise transformation}
    \vspace{-3mm}
    \label{fig3}
\end{figure*}

\subsection{Image Transformation}
First, we address a block-wise image transformation method with a secret key to encrypt test images.
As shown in Fig. \ref{fig3}, the procedure of the transformation consists of three steps: block segmentation, block transformation, and block integration.
To transform an image $x \in [0, 1]^{C \times W \times H}$, we first divide $x$ into $W_\mathrm{b} \times H_\mathrm{b}$ blocks, as in 
$\left \{ B_{1 1},B_{1 2},...,B_{W_\mathrm{b} H_\mathrm{b}} \right \}$,
where $W_\mathrm{b}=\frac{W}{M}$ is the number of blocks across width $W$, $H_\mathrm{b}=\frac{H}{M}$ is the number of blocks across height $H$, and $M$ is the block size.
In this paper, we assume that the block size of the segmentation is the same as the patch size of ConvMixer.
Next, each block is flattened, and it is concatenated again to obtain a block image $x_\mathrm{b} \in [0, 1]^{W_\mathrm{b} \times H_\mathrm{b} \times p_\mathrm{b}}$, where $p_\mathrm{b} = M^2C$ is the number of pixels in each block.
Then, $x_\mathrm{b}$ is transformed to ${x'}_\mathrm{b} \in [0, 1]^{W_\mathrm{b} \times H_\mathrm{b} \times p_\mathrm{b}}$ in accordance with block transformation with key $K$.
Finally, ${x'}_\mathrm{b}$ is transformed so that it has the same $C \times H \times W$ dimensions as those of the original image $x$, and encrypted image $x' \in [0, 1]^{C \times W \times H}$ is obtained. \par
In addition, the block transformation is carried out by using the three operations shown in Fig. \ref{fig3}.
Details on each operation are given below.

\renewcommand{\thesubsubsection}{\Alph{subsubsection}}
\subsubsection{Pixel Shuffling}
\begin{enumerate}
    \setlength{\itemsep}{0cm} 
    \item Generate a random permutation vector $\boldsymbol{v} = (v_0, v_1, ..., v_k, ..., v_{k'}, ..., v_{p_{\mathrm{b}} -1})$ by using a key $K_1$, where $k, k' \in \{0, ..., p_{\mathrm{b}} - 1 \}$, and $v_k \neq v_{k'}$ if $k \neq k'$.
    \item Pixels in each block are shuffled by vector $\boldsymbol{v}$ as
    
    \begin{equation}
        x'^{(1)}_{\mathrm{b}}(w, h, k) = x_{\mathrm{b}}(w, h, v_k).
    \end{equation}
\end{enumerate}


\subsubsection{Bit Flipping}
\begin{enumerate}
    \setlength{\itemsep}{0cm}
    \item Convert every pixel value to $[0, 255]$ scale with 8 bits (i.e., multiply $x'^{(1)}_{\mathrm{b}}$ by 255).
    \item Generate a random binary vector $\boldsymbol{r} = (r_0,...,r_k,...,r_{p_{\mathrm{b}} - 1})$, $r_k \in \{ 0,1 \}$ by using a key $K_2$. To keep the transformation consistent, $r$ is distributed with $50\%$ of ``0''s and $50\%$ of ``1''s.
    \item Apply negative-positive transformation on the basis of $\boldsymbol{r}$ as
    \begin{equation}
        x'^{(2)}_{\mathrm{b}}(w, h, k) = \left \{
        \begin{aligned}
        & x'^{(1)}_{\mathrm{b}}(w,h,k) && (r_k=0) \\
        & x'^{(1)}_{\mathrm{b}}(w,h,k) \oplus (2^L - 1) &&(r_k=1)
        \end{aligned}
        \right . ,
    \end{equation}
    where $\oplus$ is an exclusive disjunction, and $L$ is the number of bits used in $x_{\mathrm{b}}(w,h,k)$.
    \item Convert every pixel value back to $[0, 1]$ scale (i.e., divide $x'^{(2)}_{\mathrm{b}}$ by 255).
\end{enumerate}

\noindent
Since $x'^{(1)}_{\mathrm{b}}(w,h,k)$ is a floating point number between 0 and 1, bit flipping can also be expressed without scaling as follows.

\begin{equation}
    x'^{(2)}_{\mathrm{b}}(w, h, k) = \left \{
    \begin{aligned}
    & x'^{(1)}_{\mathrm{b}}(w,h,k) && (r_k=0) \\
    & 1 - x'^{(1)}_{\mathrm{b}}(w,h,k) &&(r_k=1)
    \end{aligned}
    \right .
\end{equation}


\subsubsection{Normalization}
Various normalization methods are widely used to improve the training stability, optimization efficiency, and generalization ability of DNNs.
In this paper, we also use a normalization method to achieve the combined use of transformed images and models. \par
In the normalization used in this paper, a pixel ${x'}^{(2)}_\mathrm{b}(w, h, c)$ is replaced with ${x'}_\mathrm{b}(w, h, c)$ as

\begin{equation}
    \begin{aligned}
    {x'}_\mathrm{b}(w, h, c) &= \frac{{x'}^{(2)}_\mathrm{b}(w, h, c) - 1/2}{1/2} \\
    &= 2{x'}^{(2)}_\mathrm{b}(w, h, c) - 1 \\
    &= 2(1 - {x'}^{(1)}_\mathrm{b}(w, h, c)) - 1 \\
    &= 1 - {x'}^{(1)}_\mathrm{b}(w, h, c) \\
    &= - {x'}^{(2)}_\mathrm{b}(w, h, c).
    \end{aligned}
\end{equation}

\noindent
Note that ${x'}^{(2)}_\mathrm{b}(w, h, c) = 1 - {x'}^{(1)}_\mathrm{b}(w, h, c)$ is satisfied from Eq.(6).
From Eq.(7), bit flipping with normalization can be regarded as an operation that reverses the positive or negative sign of a pixel value.
This property allows us to use the model encryption that will be described later.

\begin{figure}[h]
    \centering
    \begin{minipage}{0.45\linewidth}
    \centering
    \includegraphics[bb=0 0 251 249, scale=0.3]{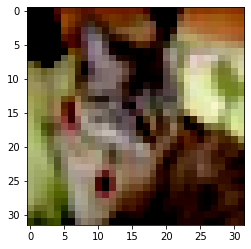}
    \subcaption{}
    \end{minipage}
    \begin{minipage}{0.45\linewidth}
    \centering
    \includegraphics[bb=0 0 251 249, scale=0.3]{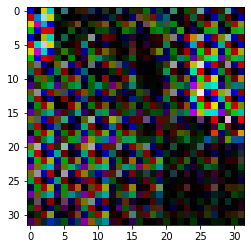}
    \subcaption{}
    \end{minipage}
    \caption{Example of images transformed by proposed method with $M = 4$. (a) Original image ($3 \times 32 \times 32$), (b) transformed image.}
    \label{fig4}
\end{figure}

\subsection{Model Transformation}
In model transformation, some parameters in models trained with plain images are transformed by using a secret key.
In this paper, a model transformation method is proposed to achieve the combined use of models and images transformed with the same key. \par
ConvMixer utilizes patch embedding (see Fig. \ref{fig1}), so it has the ability to adapt to the pixel order by patch embedding; patch embedding can be adapted to pixel shuffling and bit flipping because they can be expressed as an invertible linear transformation. \par
In the proposed method, it is assumed that the patch size $P$ used for patching is the same as the block size used for image encryption, and the number of patches is equal to that of blocks in an image.
The transformation of parameters in trained models is described below.

\subsubsection{Adaption to Pixel Shuffling}
In patch embedding, flattened patches are mapped to vectors with a dimension of $d$ as in Eq.(3).
When the patch size of ConvMixer is equal to the block size for image transformation,  $P^2 C = p_\mathrm{b}$  is satisfied.
Therefore, the permutation of rows in $\mathbf{E}$ corresponds to pixel shuffling, so the model can be encrypted with key $K_1$ used for pixel shuffling.
The accuracy of the transformed model is high only when test images are encrypted by using pixel shuffling with key $K_1$.
A permutation matrix $\mathbf{E}_1$ is defined with key $K_1$, and the transformation from matrix $\mathbf{E}$ to $\mathbf{E'}$ is shown as follows.
\begin{equation}
    \mathbf{E'} = \mathbf{E}_1 \mathbf{E}
\end{equation}

\subsubsection{Adaption to Bit Flipping}
In addition, as shown in Eq.(7), bit flipping with normalization can be regarded as an operation that randomly inverses the positive/negative sign of a pixel value.
Therefore, we can encrypt a model by inverting the sign of the rows in matrix $\mathbf{E}$ with key $K_2$ used for bit flipping.
The transformed model offers a high accuracy only for test images transformed by bit flipping with key $K_2$.
Using key $K_2$ to generate the same vector $\boldsymbol{r}$ used in bit flipping, the transformation from $\mathbf{E}$ to $\mathbf{E'}$ can be expressed as follows.
\begin{equation}
    \mathbf{E}'(k, :) = \left \{
    \begin{aligned}
    & \mathbf{E}(k, :) && (r_k=0) \\
    & - \mathbf{E}(k, :) && (r_k=1)
    \end{aligned}
    \right . ,
\end{equation}
where $\mathbf{E}(k, :)$ and $\mathbf{E}'(k, :)$ are the $k$ th rows of matrices $\mathbf{E}$ and $\mathbf{E}'$.\par

\section{Experiment and Discussion}
In an experiment, the effectiveness of the proposed method is shown in terms of image classification accuracy and model protection performance.

\subsection{Experiment Setup}
To confirm the effectiveness of the proposed method, we evaluated the accuracy of an image classification task on the CIFAR-10 dataset (with 10 classes).
CIFAR-10 consists of 60,000 color images (dimension of $3 \times 32 \times 32$), where 50,000 images are for training, 10,000 for testing, and each class contains 6,000 images.
Images in the dataset were transformed by the proposed encryption algorithm, where the block size was $4 \times 4$. \par
We used the PyTorch \cite{paszke2019pytorch} implementation of ConvMixer, where the patch size was $4$, the number of channels after patch embedding was $d=256$, the kernel size of depthwise convolution was $9$, and the number of ConvMixer layers was $8$.
The ConvMixer model was trained for $200$ epochs with Adam, where the learning rate was $0.001$.

\subsection{Image Classification}
First, we evaluated the proposed method in terms of the accuracy of image classification under the use of ConvMixer.
Table 1 shows the classification result of ConvMixer.
``Proposed'' means that ConvMixer models and test images were transformed by the proposed method.
As shown in Table 1, the proposed method did not degrade the performance at all for the transformed images. 
In contrast, the performance for plain images was degraded.
Therefore, the proposed method was effective for model protection.

\begin{table}[h]
    \centering
    \caption{Robustness against use of plain images}
    \begin{tabular}{c|cc}
    \hline
    & \multicolumn{2}{c}{Test Image} \\
    \cline{2-3}
    Model & Plain & Proposed \\
    \hline
    baseline & \textbf{90.46} & - \\
    Proposed & 11.41 & \textbf{90.46} \\
    \hline
    \end{tabular}
    \label{tab1}
\end{table}

\subsection{Model Protection}
\begin{figure}[t]
    \centering
    \includegraphics[scale=0.45]{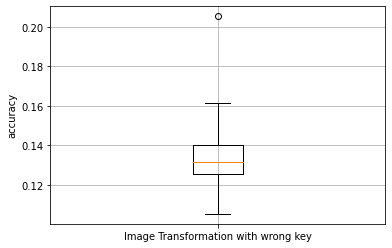}
    \caption{Evaluating robustness against random key attack. Boxes span from first to third quartile, referred to as $Q1$ and $Q3$, and whiskers show maximum and minimum values in range of $[Q1 - 1.5(Q3 - Q1), Q3 + 1.5(Q3 - Q1)]$. Band inside box indicates median. Outliers are indicated as dots.}
    \label{fig5}
\end{figure}

Next, we confirmed the performance of images encrypted with a different key from that used in model encryption.
We prepared 100 random keys, and test images encrypted with the keys were input to the encrypted model.
From the box plot in Fig. \ref{fig5}, the accuracy of the models was not high under the use of wrong keys.
Accordingly, the proposed method was confirmed to be robust against a random key attack. \par
The use of a large key spaces enhances robustness against various attacks in general.
In this experiment, the key space of pixel shuffling and bit flipping ($O_\mathsf{p}$ and $O_\mathsf{b}$) are given by $O_{\mathrm{p}} = p_{\mathrm{b}}!$ and $O_{\mathrm{b}} = \frac{p_{\mathrm{b}}!}{(p_{\mathrm{b}}/2)! \cdot (p_{\mathrm{b}}/2)!}$.
Therefore, the key space of the proposed method is $O = O_{\mathrm{p}} \times O_{\mathrm{b}} \simeq 2^{543.8}$.
The key space $O$ is sufficiently large, so it is difficult to find the correct key by random key estimation.

\section{Conclusion}
In this paper, we proposed the combined use of an image transformation method with a secret key and ConvMixer models transformed with a key.
The proposed method enables us not only to use visually protected images but to also maintain the same classification accuracy as that of models trained with plain images.
In addition, in an experiment, the proposed method was demonstrated to be robust against a random key attack.

\section*{Acknowledgments}
This study was partially supported by JSPS KAKENHI (Grant Number $\mathrm{JP}\mathrm{21} \mathrm{H}\mathrm{01327}$) and JST CREST (Grant Number $\mathrm{JPMJCR} \mathrm{20} \mathrm{D} \mathrm{3}$).











\bibliographystyle{ieice}
\bibliography{main}



\end{document}